\documentclass{jpsj2}
\title{Role of the target orientation angle and orbital
angular momentum in the evaporation residue production}

\author{Giovanni \textsc{Fazio},
Giorgio \textsc{Giardina}
\thanks{E-mail address: giardina@nucleo.unime.it},
 Francis \textsc{Hanappe}$^{1}$, Giuseppe \textsc{Mandaglio},
 Marina \textsc{Manganaro}, Akhtam I. \textsc{Muminov}$^{2}$,
 Avazbek K. \textsc{Nasirov}$^{2,3}$, and Carmelo \textsc{Sacc\'a}$^{4}$  }

\inst{INFN, Sezione di Catania,
and Dipartimento di Fisica dell'Universit{\`a} di Messina,
Messina, Italy \\
$^{1}$ Universit\'e  Libre de Bruxelles, Bruxelles, Belgium \\
$^{2}$ Department of Heavy Ion Physics, Institute of  Nuclear Physics,
Tashkent, Uzbekistan\\
$^{3}$ Joint Institute for Nuclear Research, Dubna, Russia \\
$^{4}$ Dipartimento di Scienze della Terra dell'Universit\`a di Messina, Messina, Italy}

\abst{The influence of the orientation angles of the target
nucleus symmetry axis relative to the beam direction on the production of the
evaporation residues is investigated for the $^{48}$Ca+$^{154}$Sm reaction
as a function of the beam energy.
At low energies ($E_{\rm c.m.}<$137 MeV), the yield of evaporation
residues is observed only for collisions with small
orientation angles ($\alpha_T<45^0$).
 At  large  energies (about $E_{\rm c.m.}=$140--180 MeV) all the
 orientation angles $\alpha_T$ can contribute to the evaporation residue cross section $\sigma_{ER}$ in the 10--100 mb range, and at $E_{c.m.}>$180 MeV $\sigma_{ER}$ ranges around 0.1--10 mb because the fission barrier for a compound nucleus decreases by increasing its excitation energy and angular momentum.}

\kword{Fusion, quasifission, evaporation residues, orientation of reactants}

\begin{document}
\maketitle

\section{Introduction} %% No sections necessary for express letters, letters and short notes

The study of the role of the entrance channel dynamics in the formation of
the evaporation residues (ER) in reactions with massive nuclei is an
actual problem in establishing  the conditions to obtain new superheavy
elements or new isotopes far from the island of stability of chemical
elements. The main requirements to reach maximal cross sections
in the formation of the evaporation residues are as small as possible values of the
excitation energy and angular momentum of the being formed compound
nucleus with large fusion probability. In the cold fusion reactions
the main requirements have been satisfied and 1n- and 2n-reactions
(by emission of one or two neutrons from the compound nucleus) leaded
to observe events confirming the synthesis of superheavy elements
$Z$=110 (darmstadtium), 111 (roentgenium) and  112 (see Refs.
\cite{Hofmann72,Moritajpsj76}), as well as element $Z$=113 (see Ref. \cite{MoritaJPSJ73}).
The events proving the synthesis of heavier new elements $Z$=114, 115, 116, 118
were observed in the hot fusion reactions with $^{48}$Ca on the actinide
targets $^{244}$Pu, $^{243}$Am, $^{248}$Cm and $^{249}$Cf, respectively,
in which the excitation energy of the compound nucleus was more than 35 MeV
\cite{Oganes70}.
There is an opinion  that the  values of
beam energy leading to the observed maximal cross sections
correspond to the equatorial collisions of the deformed actinide targets
\cite{Nishio06} (the orientation angle of the nucleus symmetry axis to
the beam direction is $90^{\circ}$).
The results of our calculations showed that the maximal cross sections
should be observed at orientation angles less than $90^{\circ}$
because of influence of the entrance channel on the dynamics of capture
\cite{Nasirov05}.
Therefore, to investigate the evaporation residue production, it is important to
analyze the role of the entrance channel characteristics such as the beam energy,
orbital angular momentum,
orientation angles of the symmetry axes of the projectile and target nuclei
relative to the beam direction in the angular momentum distribution of
the excited compound nucleus.
Although the reaction cross section for the interaction of massive
nuclei is enough large, only very small part
($\sigma_{ER}/\sigma_{react}\sim 10^{-9}$) can belong to the expected ER events
at synthesis of superheavy
elements $Z>108$ \cite{Hofmann72,Oganes70}. In fact, the complete
fusion of two massive nuclei is in competition with the quasifission (QF)
process \cite{Adam98} (re-separation of the interacting system into two fragments
without reaching compound nucleus) during
the first stage of the reaction, while the evaporation process (leading to
the evaporation residue nuclei) competes with the fission process along the de-excitation cascade of the compound
nucleus (CN). Moreover, at the large values of angular momentum ($\ell>82$),
also the fast-fission (FF) can contribute to the formation of nuclear
fragments hindering the ER formation.

In this paper we present the results of our study on the $^{48}$Ca+$^{154}$Sm reaction showing
as the orientation angle of the symmetry axis of the target nucleus
relative to the beam direction and orbital angular momentum
affect the yields of the evaporation residues.

\section{Method}

We use the method developed in our previous papers
\cite{Giar00,GiarArAg,FazioEPJA04,FazioJPSJ} to describe the role of
the all three stages
starting from the dinuclear system (DNS) formation
at capture of the projectile by the target-nucleus, then its evolution into
a compound nucleus and the production of the evaporation residues after
emission of gamma-quanta, neutrons, protons, $\alpha$-particles. The method
allows us to determine the corresponding cross sections of capture,
complete fusion and formation of the evaporation residues. By this method
we are able to determine the angular momentum distribution of the
DNS (determined by the conditions of the entrance channel) and competition
between quasifission and complete fusion affected by the conditions of the
reaction mechanism.
Therefore, also the de-excitation chain of the compound nucleus
(characterized by the fission-evaporation competition) is affected by the reaction dynamics
\cite{fazio,Nasirov}.

\subsection{Collision of spherical nuclei}

It is worth to calculate
the cross section of evaporation residues which can be compared
with the corresponding experimental data. The probability of
formation of the evaporation residue nucleus being survived
with mass number  $A=A_{CN}-(\nu(x)+y(x)+4k(x))$ and charge
number $Z=Z_{CN}-(y(x)+2k(x))$ from the heated and rotated compound
nucleus $^{A_{CN}}Z_{_{CN}}$ after emissions of $\nu$ neutrons,
$y$ protons, $k~\alpha$-particles at the $x$th  step of
the de-excitation cascade by the formula \cite{fazio,faziolett}:
\begin{equation}
\label{evapor}
\sigma_{ER(x)}(E^*_x)=\sum_{\ell=0}^{\ell_f}(2\ell+1)
\sigma^{\ell}_{(x-1)}(E^*_x)W_{{\rm sur}(x-1)}(E^*_x,\ell),
\end{equation}
where $\sigma^{\ell}_{(x-1)}(E^*_x)$ is the partial  cross section
of the  intermediate nucleus formation at the $(x-1)$th step and
$W_{{\rm sur}(x-1)}(E^*_x,\ell)$ is the survival probability of
the $(x-1)$th intermediate nucleus against fission along the
de-excitation cascade of CN;
$\ell_f$ is the value of angular momentum $\ell$ at which
the fission barrier for a compound nucleus
disappears completely \cite{Sierk86};
$E^*_x$ is an excitation energy of
the nucleus formed at the $x$th step of the de-excitation cascade.
 It is clear that  $\sigma^{\ell}_{(0)}(E^*_0)=\sigma^{\ell}_{\rm fus}(E^*_{CN})$
 at
\begin{equation}
\label{Ecn}
E^*_{CN}=E^*_0=E_{\rm c.m.}+Q_{gg}-E_{rot},
\end{equation}
 where $E_{\rm c.m.}$, $Q_{gg}$,
and $E_{rot}$ are the collision energy in the center of mass system,
the reaction
$Q_{gg}$-value and rotational energy of the compound nucleus, respectively.
 The numbers of the being emitted neutrons, protons, $\alpha$-particles and
$\gamma$-quanta, $\nu(x)$n, $y(x)$p, $k(x)\alpha$, and
$s(x)\gamma$, respectively, are functions of the step $x$. The
emission branching ratios of these particles depend on the
excitation energy $E^*_{A}$ and angular momentum $\ell_{A}$ of the being cooled
intermediate nucleus.

The chain of the de-excitation cascade, characterized by the emission of the
above-mentioned particles, starts from the compound nucleus $^{A_{CN}}Z_{_{CN}}$.
 Its formation probability
is the partial cross section of complete fusion
$\sigma^{\ell}_{\rm fus}(E^*_{CN})$ corresponding to the orbital
angular momentum $\ell$. The fusion cross section
is equal to the capture cross section for the light systems or light projectile
induced reactions while for the reactions
with massive nuclei it becomes a model dependent quantity. Concerning the estimation
of the fusion cross section from the experimental data of fragments, sometimes,
its value is an ambiguous
quantity because of difficulties in separation of the fusion-fission fragments from
the quasifission fragments in the case of overlap of their mass and angular distributions.

In our model, we calculate $\sigma^{\ell}_{\rm fus}(E^*_{CN})$ by estimation
of the competition of the complete fusion with quasifission if we know the
partial capture cross section:
\begin{eqnarray}
\label{s_fus} \sigma^{\ell}_{\rm fus}(E_{\rm c.m.})=
\sigma^{\ell}_{cap}(E_{\rm c.m.}) P_{\rm CN}(E_{\rm c.m.},\ell),
\end{eqnarray}
where $P_{CN}(E_{\rm c.m.})$ is the hindrance factor for the formation of the
compound  nucleus connected with the  competition between complete fusion
and quasifission as possible channels of evolution of the DNS.
Note $E_{\rm c.m.}$ and $E^*_{CN}$ are connected by relation (\ref{Ecn}).
Details of the calculation method are described in ref.\cite{faziolett}.

The partial capture cross section at given energy $E_{\rm c.m.}$ and
orbital angular momentum $\ell$ is determined by the formula:
 \begin{equation}
 \label{parcap}
 \sigma^{\ell}_{cap}(E_{\rm c.m.})=
 \pi{\lambda\hspace*{-0.23cm}-}^2
{\cal P}_{cap}^{\ell}(E_{\rm c.m.})
 \end{equation}
where ${\cal P}^{\ell}_{cap}(E_{\rm c.m.})$  is the capture
probability for the colliding nuclei to be trapped into the well of the
nucleus-nucleus potential after dissipation of a  part of the initial
relative kinetic energy and orbital angular momentum. The capture probability
${\cal P}_{cap}^{\ell}$  is equal to 1 or 0 for given $E_{\rm c.m.}$ energy and
orbital angular momentum $\ell$. Our calculations showed that in dependence
on the center-of-mass system energy $E_{\rm c.m.}$
there is a ``window'' of the orbital angular momentum for
capture with respect to the following conditions \cite{Nasirov05,FazioPRC72}:
  \[{\cal P}^{\ell}_{cap}(E_{\rm c.m.}) = \left\{
  \begin{array}{ll} 1,  \hspace*{0.2 cm}  \rm{if}
\ \  \ell_{min} \leq \ell \leq \ell_d \ \
\rm{and} \ \ {\it E_{\rm c.m.}>V}_{Coul}
  \\ 0, \hspace*{0.2cm} \rm {if}\ \ \ell<\ell_{min}   \ \
  or \ \  \ell>\ell_d  \  \rm {and} \ \
  {\it E_{\rm c.m.}>V}_{Coul}
  \\ 0,  \hspace*{0.2cm} \rm{for \  all} \ \ell \ \
     \hspace*{0.2cm} \rm{if} \ \
  {\it E_{\rm c.m.}\leq V}_{Coul}\:.
 \end{array}
 \right.
 \]

The boundary values $\ell_{min}$ and $\ell_d$ of the partial waves
leading to capture depend on the dynamics of collision and they
are determined by solving the equations of motion for the relative distance
$R$ and orbital angular momentum $\ell$ \cite{Giar00,FazioEPJA04,FazioJPSJ}.
At lower energies  $\ell_{min}$ goes down to zero and we don't
observe the $\ell$ ``window'': $0\le\ell\le\ell_d$.
The range of the $\ell$ ``window'' is defined by the size of the
potential well of the nucleus-nucleus
potential $V(R,Z_1,Z_2)$ and the values of the radial $\gamma_R$ and
tangential $\gamma_t$ friction coefficients, as well as by the moment
of inertia for the relative motion\cite{Giar00,Nasirov05}.
The capture cross section is determined  by the number of partial waves
that lead colliding nuclei to trap into the  well of
the nucleus-nucleus potential  after  dissipation of the  sufficient part
 of the initial  kinetic energy (see for example  Fig.~1(a) of Ref.
 \cite{FazioPRC72}). The size of the potential well decreases by increasing
the orbital angular momentum, $\ell$.  The  value of  $\ell$ at which
the potential well disappears is defined  as the critical value
$\ell_{cr}$. In some models, it is assumed as the maximum
value of the partial waves giving contribution to the complete fusion.
But, unfortunately, this is not true: the use of $\ell_{cr}$, as a
maximum value of $\ell$ contributing to capture,  leads to the
overestimation of the capture and fusion cross sections. Because at
$\ell_d<\ell\le\ell_{cr}$ the deep inelastic collisions take place
(see Fig.~1(b) of Ref. \cite{FazioPRC72}). It should be stressed that such
a process  occurs because of the limited values of the radial friction
coefficient \cite{Giar00,GiarArAg,PRC56fric}, the capture
becomes impossible at the low values of the orbital angular
momentum if the beam energy values are enough high than the
Coulomb barrier.

\subsection{Collision of deformed nuclei}

 Due to the dependence of the nucleus-nucleus potential ($V$) and moment
 of inertia ($J_R$)  for DNS on the orientations  of  the   symmetry axes
 of deformed  nuclei,
 the excitation functions  of  the  capture  and  fusion  are  sensitive
 to the orientations under discussion. This was demonstrated in Ref. \cite{Nasirov05}.
 The present paper is devoted  to the study of the dependence of the evaporation residue
 cross section on the orientation angles  of  the deformed interacting nuclei.
 Certainly, it is impossible directly to establish the above-mentioned dependence
 by an experimental way. But the theoretical analysis allows us to estimate
 the contributions of collisions by different orientation angles to
 the measured evaporation residue cross sections. Conclusions of such kind of
 analysis are useful to find favourable beam energies for the synthesis of
 superheavy elements in reactions with deformed nuclei.

 Usually,  the final results of the evaporation residue cross sections
  are  obtained  by averaging the contributions
 calculated for the different  orientation angles of the symmetry axes
 of the deformed reacting nuclei (as used in Ref. \cite{Nas2007})
 \begin{equation}
 \label{totalER}
<\sigma_{\small ER}(E_{\rm c.m.})>=
\int_0^{\pi/2}\sin\alpha_P\int_0^{\pi/2}\sigma_{ER}
(E_{\rm c.m.}; \alpha_P,\alpha_T))\sin\alpha_T d\alpha_Pd\alpha_T
 \end{equation}
where $\sigma_{ER}(E_{\rm c.m.};\alpha_P, \alpha_T)$ is calculated by the
formula (\ref{evapor}) for the all considered orientation angles of the
symmetry axes of the projectile and target nuclei.

\subsection{Including surface vibration of spherical nucleus}

The projectile used in the reaction under consideration is  the double
magic spherical nucleus $^{48}$Ca. But our results did not describe
experimental data at low energies if we use spherical shape for $^{48}$Ca.
Therefore, we take into account the fluctuation of its shape around the
spherical shape due to the zero-point motion connecting by the quadrupole and
octupole excitations. We calculated capture and fusion cross sections
with different vibrational states of $^{48}$Ca
$\beta_{\lambda}=-\beta^{(0)}_{\lambda},-\beta^{(0)}_{\lambda}
+\Delta\beta,...,\beta^{(0)}_{\lambda}$, where $\lambda=2,3$.
Then we performed averaging of the capture and fusion cross sections
over the values of the shape parameters used in calculations:
\begin{equation}
\label{vibaver}
<\sigma_{i}(E_{\rm c.m.},\alpha_T)>=
\int_{-\beta^{(0)}_{2}}^{\beta^{(0)}_{2}}
\int_{-\beta^{(0)}_{3}}^{\beta^{(0)}_{3}}\sigma_{i}
(E_{\rm c.m.}; \beta^{(P)}_2,\beta^{(P)}_3,\alpha_T)
g(\beta^{(P)}_2,\beta^{(P)}_3) d\beta^{(P)}_2 d\beta^{(P)}_3,
\end{equation}
with $i=cap,\, fus$ and with the weight function\cite{Esbensen81}
\begin{equation}
g(\beta^{(P)}_2,\beta^{(P)}_3)=\exp\left[-\frac{(\sum_{\lambda}\beta^{(P)}_{\lambda}Y^*_{\lambda0}
(\alpha_P))^2}{2{\sigma_{\beta_P}}^2}\right]\left(2\pi \sigma_{\beta_P}^2\right)^{-1/2}
\end{equation}
where $\beta^{(P)}_{\lambda}$ is a current value of the deformation
parameters characterizing the shape of the nucleus; for simplicity hereafter
we use $\beta_P=\{\beta^{(P)}_2,\beta^{(P)}_3\}$ characterizing the parameters
of the first collective vibrational states  2$^+$ and  3$^-$, respectively;
 $\alpha_P$ is the direction of the symmetry axis
of the projectile shape when it has prolate ($\beta^{(P)}_2>0$)
or oblate ($\beta^{(P)}_2<0$) deformation.
It is assumed  $\alpha_P=0$ in our calculations.
The dispersion of the shape fluctuations is calculated by the formula
\begin{equation}
\sigma_{\beta_P}^2=\sum_{n\lambda}\frac{2\lambda+1}{4\pi}
\frac{\hbar}{2D_{n\lambda}\omega_{n\lambda}}.
\end{equation}
for the case $n=1$ and $\lambda=2$ in $^{48}$Ca.

The amplitudes $\beta^{(0)}_{2}$, $\beta^{(0)}_{3}$,  excitation energies of the
first vibrational states $2^+$ and $3^-$ are taken from Refs.\cite{Raman,Spear},
respectively; $D_{\lambda}=\hbar/(2\omega_{\lambda}(\beta_{\lambda})^2)$.

The results obtained by (\ref{vibaver}) were used in the following formula
 \begin{equation}
 \label{totalER2}
<\sigma_{\small ER}(E_{\rm c.m.})>=
\int_0^{\pi/2}\sigma_{ER}
(E_{\rm c.m.}; \alpha_T)\sin\alpha_T d\alpha_T
 \end{equation}
to calculate the evaporation residue cross section by averaging only on the different orientation angles of the symmetry axis $\alpha_T$ of the deformed target nucleus.
The fusion excitation function is  determined by product of the partial
capture cross section $\sigma^{\ell}_{cap}$ and fusion probability
 $P_{CN}$ of DNS at various $E_{\rm c.m.}$ values:

 \begin{equation}
 \label{totfus}
 \sigma_{fus}(E_{\rm c.m.};\beta_P, \alpha_T)=\sum_{\ell=0}^{\ell_f}(2\ell+1)
 \sigma_{cap}(E_{\rm c.m.},\ell;\beta_P, \alpha_T)
 P_{CN}(E_{\rm c.m.},\ell; \beta_P, \alpha_T).
 \end{equation}

 Obviously, the quasifission cross section is defined by
\begin{equation}
 \label{totqfis}
 \sigma_{qfis}(E_{\rm c.m.}; \beta_P, \alpha_T)=
 \sum_{\ell=0}^{\ell_d}(2\ell+1)\sigma_{cap}(E_{\rm c.m.},\ell; \beta_P, \alpha_T)
 (1-P_{CN}(E_{\rm c.m.},\ell; \beta_P, \alpha_T)).
 \end{equation}

Another binary process which leads to the formation of two fragments
similar to fragments of fusion-fission or quasifission is the fast-fission.
It occurs only at high values of the angular momentum
$\ell > \ell_f$ at which the rotating nucleus has not the fission barrier
and becomes unstable against fission \cite{Sierk86}.
 It is a disintegration into two fragments of the fast rotating mononucleus which has
 survived against quasifission (the decay of the
DNS into two fragments without formation of the compound nucleus).
Therefore, the mononucleus having high values of the angular momentum,
 splits into two fragments immediately if its angular momentum is larger than $\ell_f$,
because there is not a barrier providing stability.
The fast-fission cross section  is calculated
by summing of contributions of the partial waves corresponding to the
range $\ell_f\le\ell\le\ell_d$ leading to the formation of the mononucleus:
 \begin{equation}
 \label{fasfiss}
 \sigma_{fast-fis}(E_{\rm c.m.};\beta_P,\alpha_T)=\sum_{\ell_f}^{\ell_d}(2\ell+1)
 \sigma_{cap}(E_{\rm c.m.},\ell; \beta_P, \alpha_T)
 P_{CN}(E_{\rm c.m.},\ell; \beta_P, \alpha_T).
 \end{equation}

The capture cross section in the framework of the DNS model
is equal to the sum of the quasifission,
fusion-fission and fast-fission cross sections:
 \begin{equation}
 \label{capture}
 \sigma^{\ell}_{ cap}(E_{\rm c.m.};\beta_P, \alpha_T)=
 \sigma^{\ell}_{ qfiss}(E_{\rm c.m.};\beta_P, \alpha_T)
 +\sigma^{\ell}_{ fus}(E_{\rm c.m.}; \beta_P, \alpha_T)
 + \sigma^{\ell}_{ fast-fis}(E_{\rm c.m.}; \beta_P, \alpha_T).
 \end{equation}
It is clear that fusion cross section includes the cross sections of evaporation residues and fusion-fission products.
The fission cross section is calculated by the advanced statistical
code  \cite{faziolett,ASM,dar92} that takes into account the damping
of the shell correction in the fission barrier as a function of the
nuclear temperature and orbital angular momentum.

\section{Results of the $^{48}$Ca+$^{154}$Sm reaction}

In order to investigate the influence of the angular
orientations of the interacting nuclei on the evaporation
 residue yields, we choose the $^{48}$Ca+$^{154}$Sm reaction
because the experimental data of the evaporation residue cross sections
for this reaction are presented in Ref. \cite{Stefanini05}.
Therefore, a good description of the measured ER cross section in the framework
of our model taking into account the three stages of the fusion-fission
reactions allows us to describe in detail the preceding mechanism leading
to the formation of the ER nuclei.

We study the dependence of the competition between quasifission
and complete fusion on the orientation angle  $\alpha_T$
of the symmetry axis of the target nucleus.
The quadrupole deformation parameter of $^{154}$Sm is equal to 0.27
at the ground state. Although $^{48}$Ca is spherical, in our calculations
we take into account the first quadrupole ($2^+$) and octupole ($3^-$)
collective excitations as fluctuations around the spherical shape
  with the amplitudes  $<\beta_2^{(+)}>=0.101$
 (from Ref. \cite{Raman})  and $<\beta_3^{(-)}>$=0.25 (from Ref. \cite{Spear}),
 respectively.

\begin{figure}[ht]
\begin{center}
\vspace{3.3cm}
\resizebox{0.8\textwidth}{!}{\includegraphics{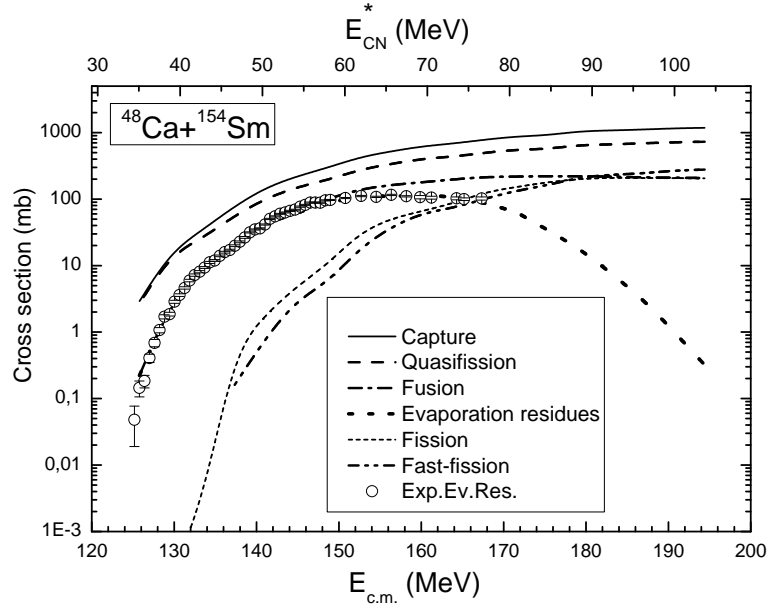}}
\end{center}
\vspace{-5.0cm}
\caption{Theoretical excitation functions for capture (solid line),quasifission (dashed line), fusion (dot-dashed line),
evaporation residue (thick short dashed line), fission (thin short-dashed
line) and fast-fission (dash-double dotted line) versus the collision energy
$E_{\rm c.m.}$ in the center-of-mass system for the $^{48}$Ca+$^{154}$Sm
reaction leading to the $^{202}$Pb compound nucleus.
Open circles are the experimental data of the evaporation residues taken
from Ref. \cite{Stefanini05}.}
\label{cross}
\vspace{-0.5 cm}
\end{figure}

 Fig. \ref{cross} shows the capture, quasifission, fusion and
evaporation residue cross sections. The complete fusion cross section
includes only the angular momentum values $\ell<\ell_f$ (see in forward
Fig. \ref{FusLorien}) because the compound nucleus becomes unstable
against fission for $\ell>\ell_f$ (where $\ell_f=82\hbar$ for $^{202}$Pb)
and appears the fast-fission. In this figure we also present
the fission and fast-fission cross sections that become appreciable
at higher values of $E_{\rm c.m.}$.
The theoretical results for the formation of evaporation residues were obtained
by using formula (\ref{totalER2}).
The good agreement between our results for ER and the
experimental data of Ref. \cite{Stefanini05} is reached by using
the formulae for the effective radii of the proton ($R_p$) and neutron
($R_n$) distributions\cite{Warda} in the nucleus as a function of
the their atomic $A$ and charge $Z$ numbers
to calculate the nucleus radius:
\begin{eqnarray}
\label{radius}
       R_{\,nucleus}(A,Z)&=&k \sqrt{(Z R_p^2  + (A-Z)R_n^2)/A},
\end{eqnarray}
where
\begin{eqnarray}
\label{radius2}
       R_p(A,Z)&=& 1.237 (1-0.157 (A-2 Z)/A-0.646/A) A^{1/3},\\
       R_n(A,Z)&=& 1.176 (1+0.25 (A-2 Z)/A+2.806/A) A^{1/3}.
\end{eqnarray}
 In our calculations we used the coefficient $k$=0.917.
 The capture and fusion cross sections depend on the radius parameter
 due to the nucleus-nucleus interaction  potential $V(R)$ between reactants.
  $V(R)$ is calculated  by the double folding procedure over the
  nucleon  distribution functions of the interacting nuclei with the effective
  Migdal's nucleon-nucleon  forces \cite{Migdal,Giar00}.
   The nucleon distribution function
    of the target-nucleus  is as follows \cite{Giar00,FazioJPSJ}:
\begin{eqnarray}
\label{rhotarg} \rho^{(0)}_T({\bf r};A,Z)&=&\rho_0
\biggl\{1+\exp\biggl[\frac{r-\tilde
R_T(\beta_2,\beta_3; \alpha_T,A,Z)}{a}\biggr]
\biggr\}^{-1},  \\
\tilde R_T(\beta_2,\beta_3; A,Z,
\alpha_T)&=&R_T(A,Z)\left(1+\beta_2 Y_{20}
(\alpha_T)+\beta_3Y_{30}(\alpha_T)\right),\nonumber
\end{eqnarray}
where $\rho_0$=0.17 fm$^{-3}$ and $a$=0.54 fm.
As we stressed above in the subsection 2.1,
 the capture probability for the given values of $E_{\rm c.m.}$ and
 $\ell$ is determined by the size of the potential  well of $V(R)$.
 The depth of the latter is used as the quasifission barrier $B_{qf}$
 in our calculations of the  fusion and quasifission cross sections
 \cite{Giar00,FazioEPJA04,FazioJPSJ,fazio}.  The fusion
 cross section depends also on the intrinsic fusion barrier $B^*_{fus}$
 for the given mass asymmetry of dinuclear system. The definition of
 $B^*_{fus}$ and its dependence on the orientation angle $\alpha_T$
 will be  discussed later
 (see Fig. \ref{drivorien}).  The values of  $B^*_{fus}$ is
 determined from the landscape of the potential energy  surface which
 includes $V(R)$. This short comment is to explain the role of
 nuclear radius in calculations of the evaporation residues
 cross section obtained by the relation (\ref{evapor}).

The ER data are comparable with the fusion cross section at $E_{\rm c.m.}$ energies
lower than 150 MeV while at the energies $E_{\rm c.m.}>$165 MeV
the fission cross section overcomes the one of the total evaporation residues
 becoming comparable with the fusion cross section (see Fig. \ref{cross}).
Also the fast-fission cross section becomes appreciable at energies
$E_{\rm c.m.}>$165 MeV  while the evaporation
residue cross section decreases.
Such a decrease of $\sigma_{ER}$
is connected with the fact that the fission barrier for a compound nucleus
decreases by increasing its excitation energy E$^*_{CN}$\cite{FazioEPJA04,FazioJPSJ}
and angular momentum $\ell$\cite{Sierk86}. Therefore, the survival probability of the heated and
rotating nuclei along the de-excitation cascade of CN strongly decreases.

\begin{figure}[h]
\begin{center}
\vspace{2.8cm}
\resizebox{0.8\textwidth}{!}{\includegraphics{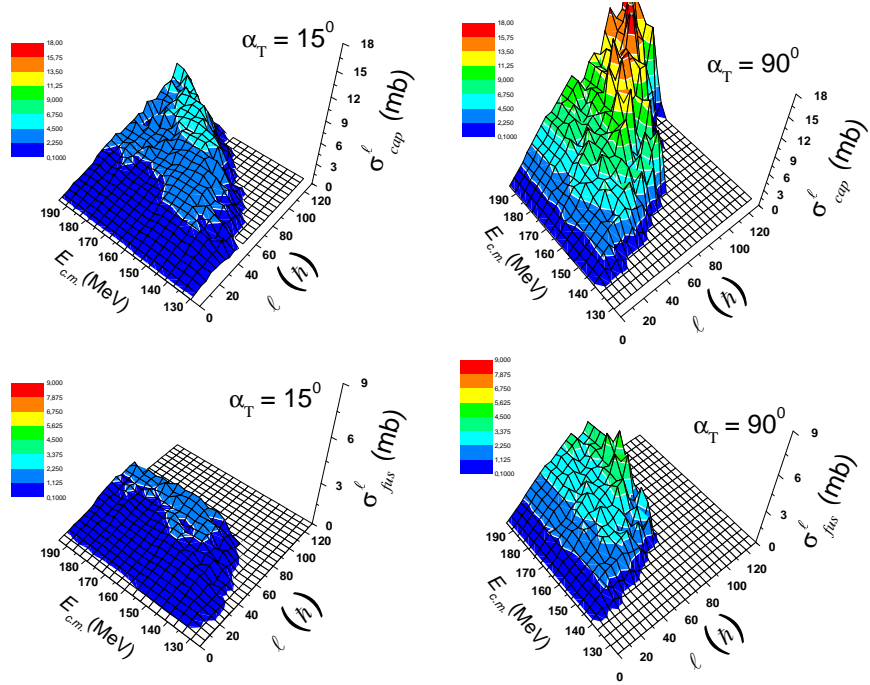}}
\end{center}
\vspace{-4.0cm}
\caption{Spin distribution of the capture (upper part) and fusion (lower part) cross sections
as a function of the collision energy $E_{\rm c.m.}$
in the center-of-mass-system  and angular momentum $\ell$, at  orientation
angles $\alpha_T=15^\circ$ and $90^\circ$. Note that the scale of $\sigma^{\ell}_{fus}$
is two times smaller than the one for $\sigma^{\ell}_{cap}$ in order to see in a better
way the structure of the $\sigma^{\ell}_{fus}$ shape.}
\vspace{-0.5cm}
\label{oriencapfus}
\end{figure}

\begin{figure}[h]
\begin{center}
\vspace{2.7cm}
\resizebox{0.85\textwidth}{!}{\includegraphics{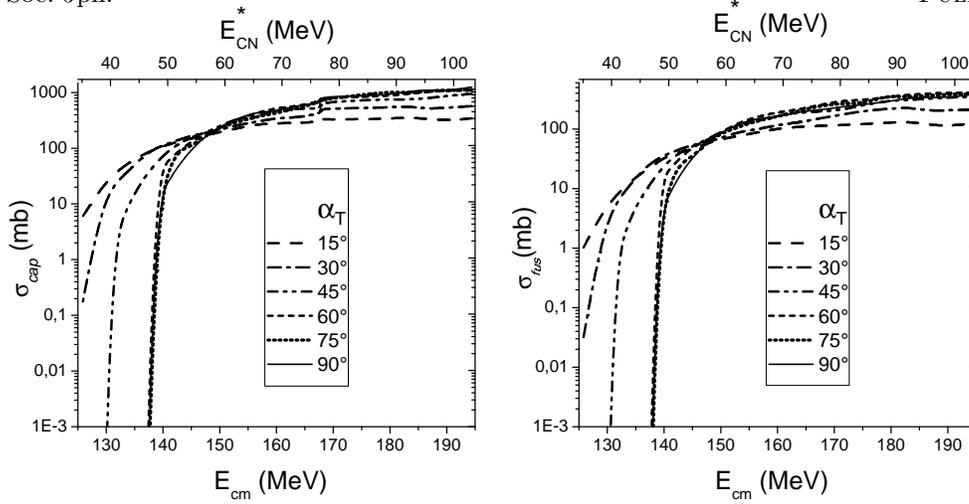}}
\end{center}
\vspace{-5.8cm}
\caption{Capture  and fusion  cross sections versus the collision energy
$E_{\rm c.m.}$ for various target
orientation angles $\alpha_T$.}
\vspace{-0.5cm}
\label{oriencap}
\end{figure}

 Before  the analysis of  the contributions of the different
orientation angles $\alpha_T$ of the symmetry axis of the $^{154}$Sm
 target  to the averaged results for the evaporation residues in Fig. \ref{cross}
 we discuss the dependence of the partial capture and fusion cross sections
 on $\alpha_T$. The presented results of the
 dependence on the orientation angle  $\alpha_T$ in Figs.
 \ref{oriencapfus}, \ref{oriencap}, \ref{FusLorien} and \ref{QfissSpin}
  are already multiplied on the weight factor $\sin\alpha_T$ used at
  averaging procedure by (\ref{totalER2}).

The upper and lower parts of Fig. \ref{oriencapfus}  show
the partial capture ($\sigma^{\ell}_{cap}$) and  fusion
($\sigma^{\ell}_{fus}$) cross sections, respectively, as a function of the energy $E_{c.m.}$ for
some typical orientation angles (for example, at $\alpha_T$=15$^\circ$ for a small angle,
and $\alpha_T$=90$^\circ$ for a large angle) of the symmetry axis of the deformed $^{154}$Sm
target nucleus.
The volumes of the distributions strongly depend on the orientation
angle $\alpha_T$ of the target.

\begin{figure}[ht]
\begin{center}
\vspace{3.5 cm}
%Coulbarr.eps
\resizebox{0.75\textwidth}{!}{\includegraphics{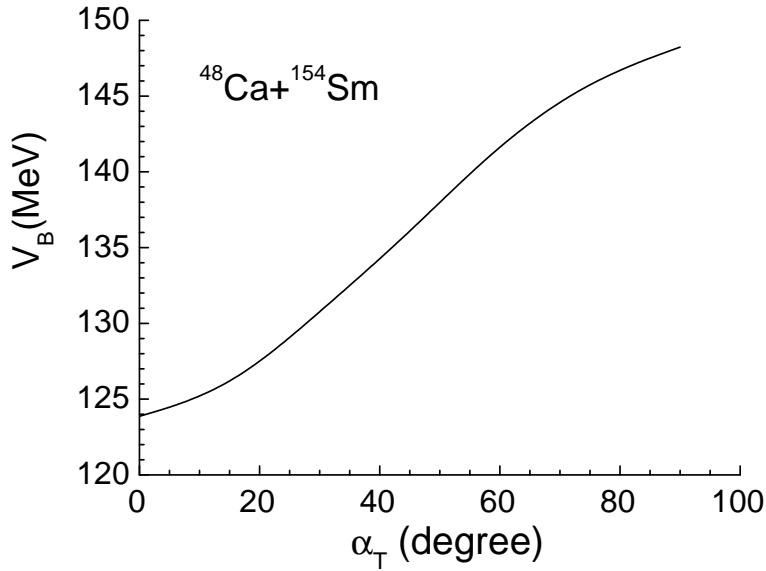}}
\end{center}
\vspace{-4.5cm}
\caption{Coulomb barrier $V$ of the nucleus-nucleus interaction versus the
orientation angle $\alpha_T$ of the target nucleus.}
\vspace{-0.5cm}
\label{Coulbarr}
\end{figure}

\begin{figure}[ht]
\begin{center}
%\vspace{3.1cm}
\resizebox{0.85\textwidth}{!}{\includegraphics{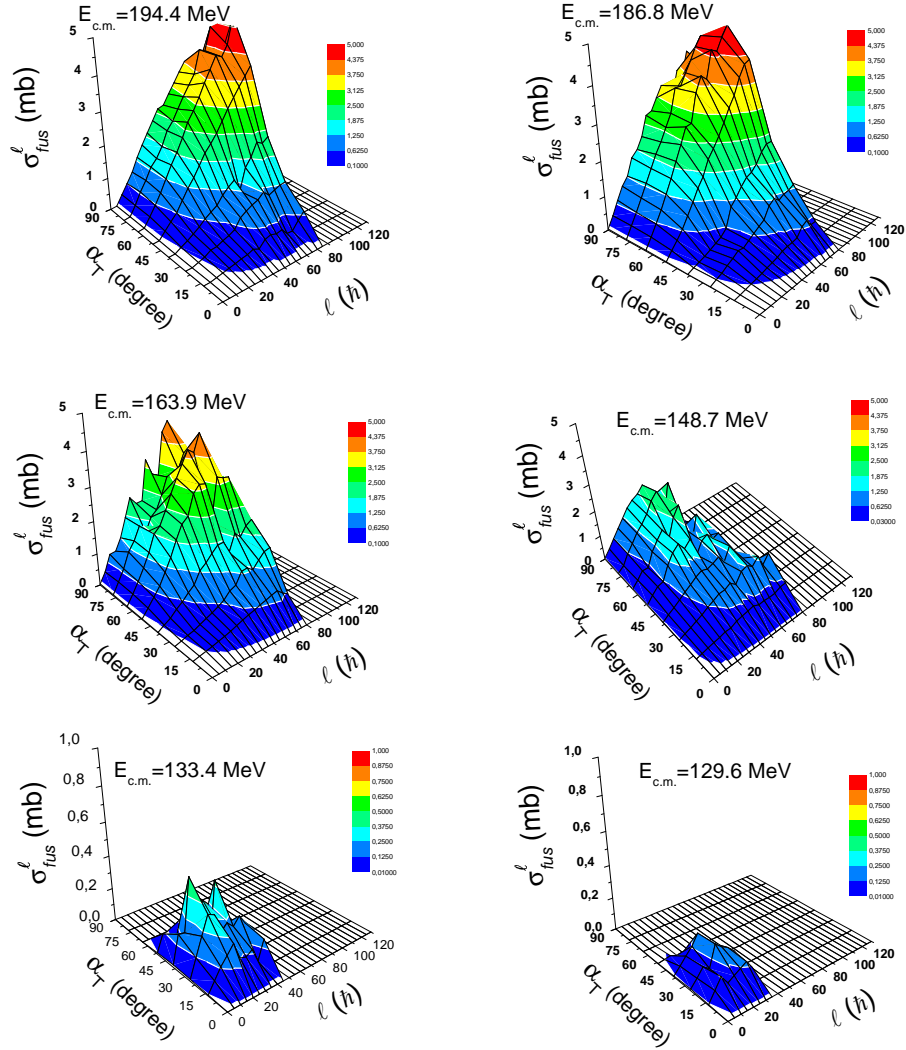}}
\end{center}
%\vspace{-4.8cm}
\vspace{-0.8cm}
\caption{Partial fusion cross section as a function of the
orientation angle $\alpha_T$ of the target nucleus and initial orbital angular
momentum $\ell$, at various values of the collision energy $E_{\rm c.m.}$. Note that the scale of the panels at $E_{\rm c.m.}=129.6$ and  133.4 MeV is five times smaller than the one of other panels in order to see in a better way the structure of the $\sigma^{\ell}_{fus}$ shape.}
\vspace{-0.5cm}
\label{FusLorien}
\end{figure}

\begin{figure}[ht]
\begin{center}
%\vspace{3.1cm}
\resizebox{0.85\textwidth}{!}{\includegraphics{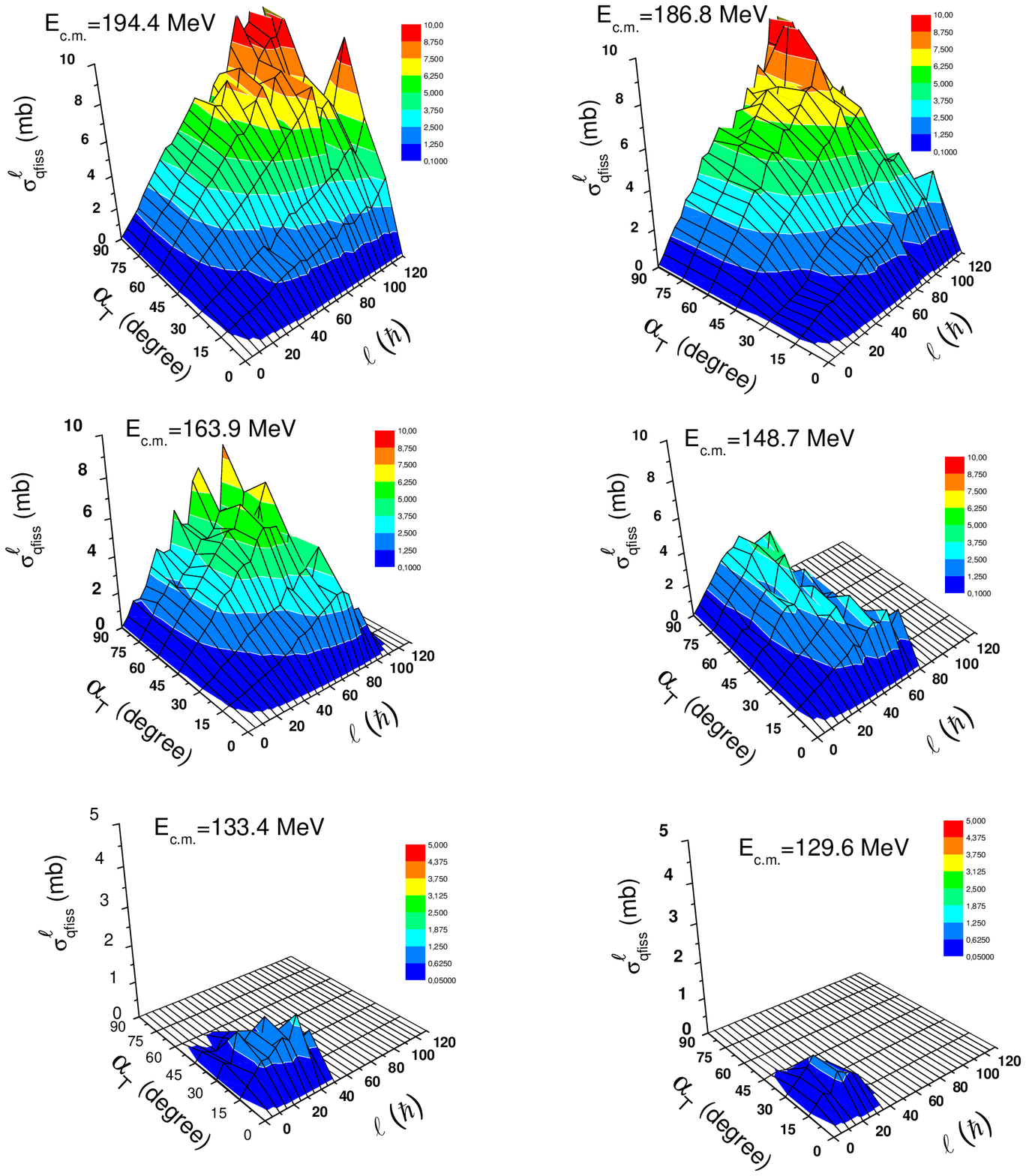}}
\end{center}
%\vspace{-4.8cm}
\vspace{-0.8cm}
\caption{Partial quasifission cross section as a function of the
orientation angle $\alpha_T$ of the target nucleus and initial orbital angular
momentum $\ell$, at various values of the collision energy  $E_{\rm c.m.}$. Note that the scale of the panels at $E_{\rm c.m.}=129.6$ and  133.4 MeV is two times smaller than the one of other panels in order to see in a better way the structure of the $\sigma^{\ell}_{qfiss}$ shape.}
\vspace{-0.5cm}
\label{QfissSpin}
\end{figure}

 From Fig. \ref{oriencapfus} appears that for the
considered $^{48}$Ca+$^{154}$Sm reaction the capture yield strongly overcomes the
fusion one at each angular configuration $\alpha_T$ because quasifission is  the dominant
 process in the competition with fusion during the evolution of the DNS, due to the
 relatively small values of the quasifission barrier $B_{qf}$ (see in forward panel
 (b) of Fig. \ref{Qfisbarr}). At low energies (about $E_{c.m.}$ = 125 - 137 MeV) the fusion,
 reached by the evolution of the DNS, is low and appears for small orientation angles
 $\alpha_T$ of the target (see low part of Fig. \ref{oriencapfus}, left panel at $\alpha_T=15^\circ$). Instead, at large values of $\alpha_T$ (see the right panel at $\alpha_T$ = 90$^\circ$) the fusion formation is not possible in the above-mentioned $E_{\rm c.m.}$ energy range.  For large orientation angles $\alpha_T$ the fusion process can occur only at larger $E_{\rm c.m.}$ energies due to the large values of the Coulomb barrier (see in forward Fig. \ref{Coulbarr}).

Fig.  \ref{oriencap} shows the capture and fusion cross section versus $E_{c.m.}$ for many different orientation angles $\alpha_T$ of the target.
This figure shows  large capture cross sections at high values of the $E_{c.m.}$ energy.
 At lower $E_{c.m.}$ energies (at about  $E_{c.m.}<$ 137 MeV) only the
 small orientation angles of the target ($\alpha_T \leq 45^{\circ}$) can give
 contribute to the capture cross section due to the low values of the
 Coulomb barrier for the mentioned $\alpha_T$ angle range (see again Fig. \ref{Coulbarr}).
At about $E_{c.m.}$=148 MeV all the $\alpha_T$ configurations contribute to the capture cross section approximately with the same possibilities because the collision energy $E_{\rm c.m.}$ is enough to overcome the maximum value of the Coulomb barrier depending by the $\alpha_T$ angle. In the above-mentioned low energy range, also for the fusion cross section can contribute only the small configuration of the target with $\alpha_T \leq 45^\circ$.  At higher $E_{\rm c.m.}$ energies (at about $E_{\rm c.m.}>$ 155 MeV) the contributions for the configurations with $\alpha_T \geq 45^{\circ}$ are larger than the ones with $\alpha_T \leq 30^{\circ}$ for both the capture and fusion cross sections.

The dependence of the Coulomb barrier on $\alpha_T$ is
presented in Fig.\ref{Coulbarr}.
This phenomenon is evident from the geometry of collision in relation to the relative distance between the centers of the two interacting nuclei.
For large orientation angles $\alpha_T$, the capture  process
can occur at high values of  $E_{\rm c.m.}$ only because increases  the
Coulomb barrier (due to the decrease of the distance of centers of nuclei by increasing the angle $\alpha_T$)  determined by the formula A.2 discussed in Ref. \cite{FazioEPJA04}. The Coulomb barrier
reaches its maximum value $V_B=$ 148.2 MeV  at $\alpha_T$= 90$^{\circ}$ (equatorial collision).
Such behaviour of the Coulomb barrier is expected.
But a dependence of the hindrance to the complete fusion on $\alpha_T$ is not evident.

In Figs. \ref{FusLorien} and \ref{QfissSpin} are presented the partial cross sections
of complete fusion $\sigma^{\ell}_{fus}$ and quasifission $\sigma^{\ell}_{qfiss}$,
respectively, as a function of the orientation angle $\alpha_T$ for different values of
the collision energy in the center-of-mass system.
The volumes of the distributions strongly depend on the values of $\alpha_T$.
The $\sigma^{\ell}_{qfiss}(E_{\rm c.m.})$ and  $\sigma^{\ell}_{fus}(E_{\rm c.m.})$
cross sections are related to the partial capture cross section
$\sigma^{\ell}_{cap}(E_{\rm c.m.})$ by the formula (\ref{capture}) which also
includes the contribution of the fast-fission cross section
$\sigma^{\ell}_{fast-fis}(E_{\rm c.m.})$.
From these figures appears that for the considered
$^{48}$Ca+$^{154}$Sm reaction the  quasifission is the dominate process
for all orientation angles  $\alpha_T$ of the target-nucleus.

The smallness of the capture and fusion
cross sections at small orientation angles $\alpha_T$ is
connected with the restriction of the number of partial waves  leading
to the capture because the entrance channel barrier increases by  increasing the
DNS rotational energy at given
values of $\alpha_T$. This effect is seen from the bottom panels of
Figs. \ref{FusLorien} and \ref{QfissSpin}.
 According to the dinuclear system model
the depth of the potential well for the given mass asymmetry of reactants is
considered as a quasifission barrier $B_{qf}$ keeping DNS from decay
into two fragments with the corresponding mass and  charge numbers.
Peculiarities of such a dependence reflects on the capture cross section
 and  competition between quasifission and complete fusion processes
 during the evolution of the dinuclear system.

 Calculations of dynamics of incoming paths show the following properties of
 the capture cross section:

 --  the capture of the projectile by the target nucleus takes place if the collision energy is larger than the Coulomb barrier for each collision with the
 corresponding orientation  angle $\alpha_T$;

 -- the number of the partial waves which determine the capture cross section
 increases by increasing the beam energy;

 -- the number of the partial waves is larger if the depth of the potential
   well is large;

 -- the number of the partial waves ceases to increase by increasing the
 beam energy for the given  orientation  angle $\alpha_T$
 if the beam energy is enough larger than the Coulomb barrier due to the
 restricted value of the radial friction coefficient. In fact, after dissipation
 of the relative kinetic energy the projectile could not be trapped
 into the potential well. So, for small values of the
 orbital angular momentum and large values of the beam energy
 we have a ``$\ell$-window'' because $\ell_{min} > 0$ (see page 4).

\begin{figure}[th]
\begin{center}
\vspace{4.1cm}
\resizebox{0.85\textwidth}{!}{\includegraphics{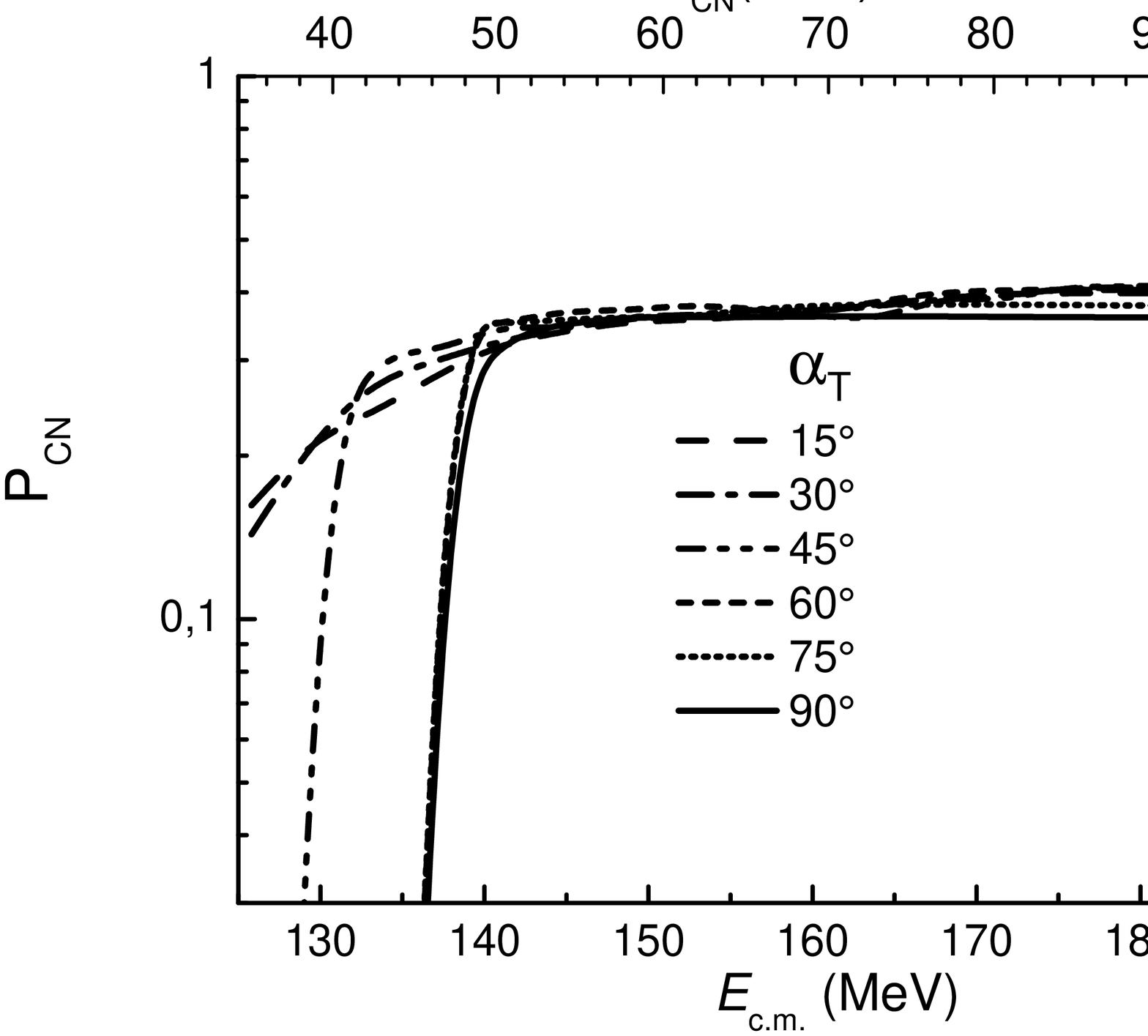}}
\end{center}
%\vspace{-4.8cm}
\vspace{-4.3cm}
\caption{Fusion probability $P_{CN}$ versus the collision energy
$E_{\rm c.m.}$ for different values of the $\alpha_T$ angle.}
\vspace{-0.9cm}
\label{pcnorien}
\end{figure}

\begin{figure}[htb]
\begin{center}
\vspace{3.0cm}
%Qfisbarr.eps
\resizebox{0.9\textwidth}{!}{\includegraphics{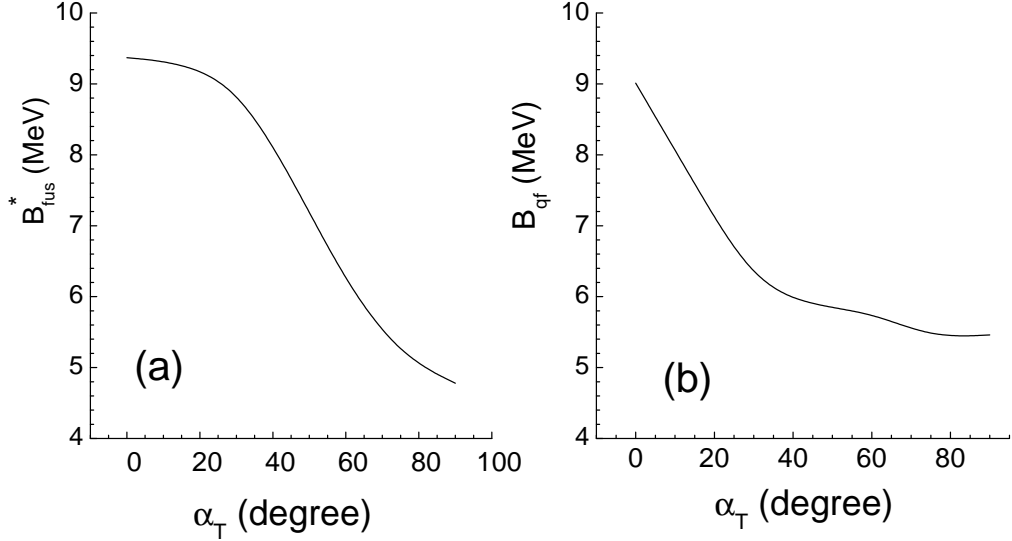}}
\end{center}
\vspace{-6.5 cm}
\caption{Intrinsic fusion barrier $B^*_{fus}$ (a) and  quasifission
barrier $B_{qf}$ (b) for the initial charge asymmetry of the
$^{48}$Ca+$^{154}$Sm reaction leading to the $^{202}$Pb compound nucleus versus
the orientation angle $\alpha_T$ of the target--nucleus.}
\vspace{-0.6cm}
\label{Qfisbarr}
\end{figure}

\begin{figure}[h]
\begin{center}
\vspace{3.5cm}
\resizebox{0.8\textwidth}{!}{\includegraphics{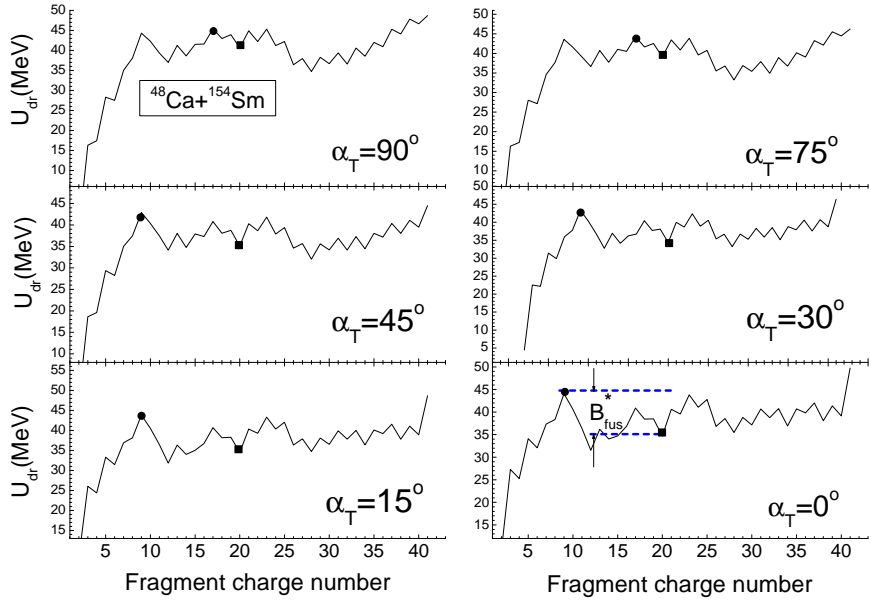}}
\end{center}
\vspace{-4.2cm}
\caption{The dependence of the driving potential $U_{dr}$ calculated
for the dinuclear system formed in the $^{48}$Ca+$^{154}$Sm reaction for different  values of the orientation angle $\alpha_T$.
 The difference between the values of $U_{dr}$  corresponding  to the solid square at the initial charge asymmetry and solid circle
 on the maximum value of the driving potential is equal to
 $B^*_{fus}$.}
\label{drivorien}
%\vspace{-0.9cm}
\end{figure}

 We stress that by increasing the
 beam energy, also increases the possible angular configurations of the
 deformed target that contribute to the fusion process. At  higher energies
 the rate of the fusion formation at competition between quasifission and fusion
  increases (see in forward Fig. \ref{pcnorien}).
  The reason is connected with the decrease of the intrinsic fusion barrier
$B^*_{fus}$ by the increase of the orientation angle  $\alpha_T$.

The increase of hindrance to the complete fusion at small orientation
angles was discussed in Ref.\cite{Nasirov05} and we observe it
clearly for the investigated $^{48}$Ca+$^{154}$Sm reaction.
This is connected with the dependence of the potential energy
surface, particularly of the driving potential $U_{dr}$, on the
orientation angle of the symmetry axis of deformed nucleus
relative to the beam direction. As a result the competition between complete
fusion and quasifission becomes a function of the angular configuration
of the reacting nuclei (see Ref. \cite{Nasirov05}). For example, the role of
the quasifission in reactions with deformed nuclei
 was discussed in the paper of Hinde {\it et al.}\cite{Hinde53}
 showing the increase of the anisotropy of the fragment angular
 distribution at lowest $E_{\rm c.m.}$ energies
for the $^{16}$O+$^{238}$U reaction.

Figs. \ref{cross} and \ref{QfissSpin} show the increase of the quasifission
cross section $\sigma^{\ell}_{qfiss}$ by increasing the collision energy
$E_{\rm c.m.}$ but the fusion cross section $\sigma^{\ell}_{fus}$
also increases by  $E_{\rm c.m.}$ more fastly than
$\sigma^{\ell}_{qfiss}$ before saturation of $P_{CN}$.
Favorable conditions for the formation of the evaporation residues in
collisions with the large orientation angles $\alpha_T$ can be
seen from the study of the fusion factor $P_{CN}$ which
depends on $\alpha_T$.
The fusion probability $P_{CN}$ is calculated by using the
  formula (11) of Ref. \cite{Nas2007}.

Fig. \ref{pcnorien} shows the fusion probability $P_{CN}$ as
 a function of the collision energy $E_{\rm c.m.}$, for different
 values of the target orientation angle $\alpha_T$.
 Also the results of $P_{CN}$ confirm the conclusion that
 at $E_{\rm c.m.}<$137 MeV  only collisions with small
  $\alpha_T$ values (about $\alpha_T\leq$ 45$^\circ$) can
  give an appreciable (small values) fusion probability $P_{CN}$.
  In such cases we obtain a reduction of the CN formation.
  By increasing the $E_{\rm c.m.}$ also the fusion probability
  increases and the contributions of collisions with large
  angles up to $\alpha_T= 90^\circ$ appear.
   At about  $E_{\rm c.m.}>$ 150 MeV the fusion probability
   $P_{CN}$ ranges around 0.35-0.41 with some saturation.

It is difficult to study quantitatively the dependence of the
quasifission cross section as a function of $E_{c.m.}$ by the experimental data only because it is difficult to separate unambiguously
quasifission fragments from the deep-inelastic collision,
fusion-fission and fast-fission fragments.

The contribution of the quasifission process in the observed
anisotropy of the fragment angular distribution was explored in Ref. \cite{Nas2007}
in the framework of the model based on the dinuclear system-concept:
competition between complete fusion and quasifission is related to both
values of the intrinsic fusion $B^*_{fus}$ and
quasifission $B_{qf}$ barriers (see for example formula
(11) in Ref.\cite{Nas2007}).

 According to our calculations increase of the yield of
 the quasifission fragments is explained by decreasing of
 the quasifission  barrier $B_{qf}$.
 The small value of $B_{qf}$ means
   less stability of the dinuclear system during its evolution.
   There are two reasons leading to the decrease of
 the  quasifission barrier:  $B_{qf}$ decreases  by increasing
 $\alpha_T$ (see  Fig. \ref{Qfisbarr}(b)) and angular momentum $\ell$.
Large values of both $\alpha_T$ and  $\ell$
 are populated by increasing $E_{\rm c.m.}$.
In Fig. \ref{Qfisbarr} we present the calculated values
of the $B^*_{fus}$ and $B_{qf}$ barriers for the $^{48}$Ca+$^{154}$Sm
reaction as a function of the orientation angle $\alpha_T$.

In the DNS model, $B^*_{fus}$
is determined as a difference of the values of the driving potential
which corresponds to the initial mass asymmetry (solid squares in
Fig. \ref{drivorien}) and its maximum value in the way to the complete
fusion along the mass asymmetry axis $Z\rightarrow 0$ (solid circles
 in Fig. \ref{drivorien}). In the case of collisions of deformed  nuclei,
the values of $B^*_{fus}$ are larger for small orientation
angles in comparison with the ones for large orientation angles of the
symmetry axes of nuclei relative to the beam direction.
In Fig. \ref{drivorien}, the driven potential is presented in
dependence on the values $\alpha_T$ for the$^{48}$Ca+$^{154}$Sm
reaction.  One can see that by the increase of the orientation angle
 $\alpha_T$  of the target leads to the decrease of the intrinsic fusion barrier
 $B^*_{fus}$ considered relative to the initial charge number $Z=20$.
 As a result the fusion probability $P_{CN}$ that appears in formula
 (\ref{s_fus}) is sensitive to both
 values of the $B^*_{fus}$ and $B_{qf}$ barriers.
 But, as we have already stressed, the capture can occur
at the large values of $\alpha_T$ only if we increase the beam energy
allowing the system to overcome the Coulomb barrier.

Such values and trends of the barriers are consistent with the results
that at lower $E_{\rm c.m.}$ energies only the small angles $\alpha_T$
can contribute to the $\sigma_{cap}$, $\sigma_{fus}$ and $\sigma_{ER}$
cross sections, by the relevant role of quasifission.
At higher $E_{\rm c.m.}$ energies also large $\alpha_T$
configurations contribute to the above-mentioned cross sections.

\begin{figure}[ht]
\begin{center}
\vspace{3.7cm}
\resizebox{0.8\textwidth}{!}{\includegraphics{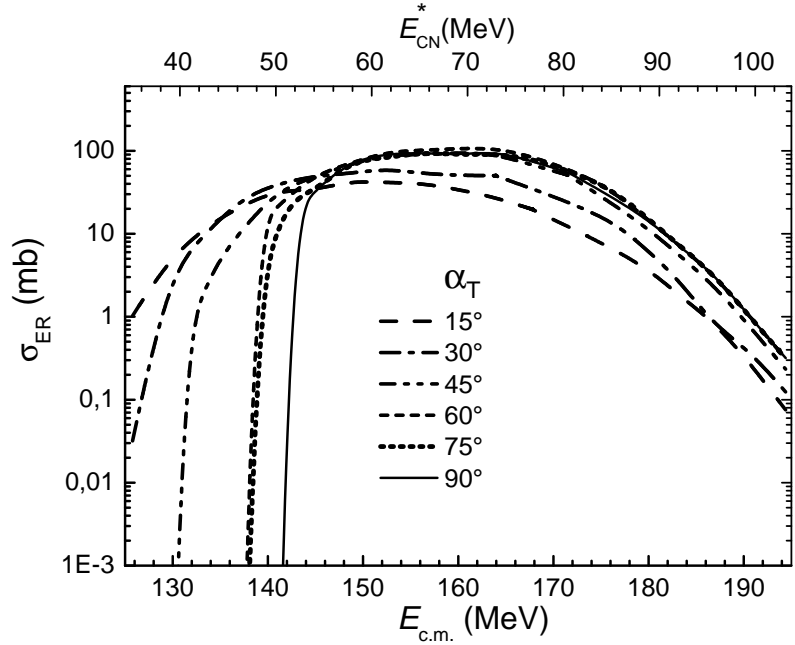}}
\end{center}
\vspace{-4.2cm}
\caption{Evaporation residue cross sections versus the
$E_{\rm c.m.}$ energy for various orientation angles $\alpha_T$
 of the $^{154}$Sm target.}
\vspace{-0.5cm}
\label{orient1}
\end{figure}

Fig. \ref{orient1} shows the excitation function of the
evaporation residue formation, at various target orientation angles.
At the smallest energy $E_{\rm c.m.}$=125.8 MeV (corresponding to $E_{CN}^*$=35.1
MeV of the compound nucleus) the main contributions to the evaporation residue
cross section are given
only by collisions with angular orientations of the target $\alpha_T<30^{\circ}$.
This is explained by the low values of the Coulomb barrier
for such orientation angles $\alpha_T$ and by the appreciable probability (see Fig. \ref{pcnorien}) of complete fusion
in competition with quasifission. In any case,
at energies  $E_{\rm c.m.}<$ 137 MeV  only
collisions with the orientation angles about $\alpha_T\leq$ 45$^{\circ}$
 can contribute to the ER formation.
At $E_{\rm c.m.}>148$ MeV, collisions with all
orientation angles contribute to the ER formation and the
$\sigma_{ER}$ cross section ranges between 10--100 mb. The total ER yield reaches the maximum value (see also Fig. \ref{cross}) with a saturation in the  148--158 MeV energy range. Then, at higher $E_{\rm c.m.}$ energies
the formation of the evaporation residues decreases and at $E_{\rm c.m.}>$180 MeV $\sigma_{ER}$ ranges around 0.1--10 mb. In spite of the approximately constant trend of the fusion probability $P_{CN}$ and fusion cross section $\sigma_{fus}$ at $E_{\rm c.m.}>$ 160 MeV, the  decrease of $\sigma_{ER}$ is explained by the increase of the excitation energy of the compound nucleus and its angular momentum which lead to a  decrease of the fission barrier $B_f$
of the compound nucleus. The angular momentum distribution of the heated and
rotated compound nucleus (and other intermediate nuclei along the
de-excitation cascade of CN) plays a decisive role in calculation of its
survive probability against fission, and then to the formation of the
evaporation residues \cite{FazioEPJA04}. Another phenomenon leading to the decrease of $\sigma_{ER}$ at higher
collision energies $E_{\rm c.m.}$ is the fast-fission process which is the splitting of the mononucleus
into two fragments due to the absence of the fission barrier at very
high values of the angular momentum  $\ell > \ell_f$.

At least, we also discuss the result when we include in calculations the tunneling effect for the collisions at subbarrier fusion energy. This effect leads to the shift of the capture excitation function on 1.5--2.0 MeV to lower energies. In mass asymmetric reactions (for example $^{16}$O+$^{238}$U) the fusion cross section is nearly equal to the capture cross section and the enhancement of the subbarrier fusion can be seen in the cross section of the evaporation residues. In reactions with massive nuclei (with a higher mass symmetric parameter, as for example the  $^{48}$Ca+$^{154}$Sm reaction), taking into account tunneling through the interaction barrier leads to an enhancement of the capture probability but this effect on the fusion excitation function for this reaction is very small.  This is due to the intrinsic fusion barrier $B^*_{fus}$ which decreases the fusion probability at a given capture cross section. $B^*_{fus}$ is about 10 MeV for collisions with small orientation angles $\alpha_T$ and it decreases up to 5 MeV for large orientation angles (see Fig. \ref{drivorien}). Due to the smallness of this effect on the results under discussion we do not present the results of such estimations.

\section{Conclusions}

In this paper the role of the orientation angle $\alpha_T$
 of the symmetry axis of the deformed $^{154}$Sm nucleus
 on  the complete
fusion and evaporation residue cross sections is studied for the
$^{48}$Ca+$^{154}$Sm reaction at energies near and above the Coulomb barrier.
The dependence of the quasifission-fusion competition during the
evolution of the dinuclear system and the sensitivity of the
fission-evaporation competition during the de-excitation cascade of the
compound nucleus on the values of the orientation angle  $\alpha_T$
are demonstrated. The analysis of the dependence of the  fusion and
evaporation residue cross sections on the $\alpha_T$  angle shows that
the observed yield of evaporation residues in the $^{48}$Ca+$^{154}$Sm reaction
at low energies ($E_{\rm c.m.}<$137 MeV) is formed at collisions with orientations $\alpha_T<45^{\circ}$ because the collision energy is enough
to overcome the corresponding Coulomb barrier. Only in this cases it is
possible  formation of the dinuclear system
which evolves to the compound nucleus or breaks up into two fragments after multinucleon
exchange without formation of the compound nucleus.
  At high energies (about in the $E_{\rm c.m.}=$148--180 MeV energy range) all
 orientation angles of the target-nucleus can contribute
to $\sigma_{ER}$ and its values are included in the 10--100 mb interval. In spite of high collision energies (about $E_{c.m.}>$158
MeV) the complete fusion still increases, the evaporation residue cross section $\sigma_{ER}$ goes
 down and its values range in the 0.1--1 mb interval due to the
 strong decrease of the survival probability against fission of the heated
 compound nucleus along de-excitation cascade. This is connected
 by the decrease of the  fission barrier for a compound nucleus
 by increasing its
excitation energy \cite{FazioEPJA04,FazioJPSJ} and angular momentum \cite{Sierk86}.

Another phenomenon leading to the decrease of $\sigma_{ER}$ at  high
 energies is the fast-fission process which is the splitting of the mononucleus
into two fragments due to absence of the fission barrier  at very high the angular
momentum  $\ell > \ell_f$ where $\ell_f$ is the value of angular momentum
at which  barrier disappears.

From our study on the $^{48}$Ca+$^{154}$Sm reaction leading to the $^{202}$Pb compound nucleus we can affirm that  in the explored energy range of collisions the quasifission is the dominant process in competition with the complete fusion for any orientation angle  $\alpha_T$ of the symmetry axis  of the deformed target--nucleus.

\section*{Acknowledgment}
Authors thank Prof. R.V. Jolos, Drs. G.G. Adamian and N.V. Antonenko for fruitful discussions of this paper.
G. Giardina and A.K. Nasirov are grateful to the Fondazione Bonino-Pulejo
(FBP) of Messina for the important  support received in the international
collaboration between the Messina group and the Joint Institute for Nuclear
Research of Dubna (Russia). One of authors A. Nasirov thanks INFN, Sezione di Catania,
and Dipartimento di Fisica dell'Universit{\`a} di Messina for warm
hospitality. This work was performed partially under the financial
support of the RFBR and INTAS.

\end{document}